\documentclass[prl,twocolumn,showpacs]{revtex4}

\newcommand{\be}{\begin{equation}}
\newcommand{\ee}{\end{equation}}
\newcommand{\bea}{\begin{eqnarray}}
\newcommand{\eea}{\end{eqnarray}}

\usepackage{graphicx}

\begin{document}
\title{Crystalline Order on a Sphere and the Generalized Thomson Problem}

\author{M. Bowick,$^{1}$ A. Cacciuto,$^{1}$ D. R. Nelson,$^{2}$ and 
A. Travesset$^{3}$}

\affiliation{$^1$ Physics Department, Syracuse University,  
Syracuse NY 13244-1130, USA } 
\affiliation{$^2$ Lyman Laboratory of Physics, Harvard University,
Cambridge MA 02138, USA} 
\affiliation{$^3$ Loomis Laboratory, University of Illinois at Urbana, Urbana IL 
61801, USA}   
 
\begin{abstract}
We attack generalized Thomson problems with a continuum formalism 
which exploits a universal long range interaction between defects
depending on the Young modulus of the underlying lattice. 
Our predictions for the ground state energy agree with simulations of 
long range power law interactions of the form $1/r^{\gamma}$ ($0 < 
\gamma < 2$) to four significant digits. The regime of grain
boundaries is studied in the context of tilted crystalline order and
the generality of our approach is illustrated with new results for
square tilings on the sphere.

\end{abstract}
\maketitle
The Thomson problem of constructing the ground state of (classical) 
electrons interacting with a repulsive Coulomb potential on a 2-sphere
\cite{Thom:04} is almost one hundred years old \cite{Sku:97} and has
many important physical realizations. These include multi-electron
bubbles \cite{Lei:93}, which may be studied by capillary wave
excitations, or the surface of liquid metal drops
confined in Paul traps \cite{Dav:97}. Although the original Thomson
problem refers to the ground state of spherical shells of electrons,
one can also ask for crystalline ground states of particles
interacting with other potentials. Such a generalized Thomson problem  
arises, for example, in determining the arrangements of the protein
subunits which comprise the shells of spherical viruses
\cite{Virus:62,Virus:93}. Here, the ``particles" are clusters of 
protein subunits arranged on a shell. Other realizations include
regular arrangements of colloidal particles in {\em colloidosomes} 
\cite{Nik:01} proposed for encapsulation of active ingredients such
as drugs, nutrients or living cells \cite{DHNMBW:02} and fullerene
patterns of carbon atoms on spheres  \cite{Kr:85} and other geometries
\cite{Rings:2001}. An example with long range (logarithmic)
interactions is provided by the Abrikosov lattice of vortices which
would form at low temperatures in a superconducting metal shell with a
large monopole at the center \cite{DoMo:97}. In practice, the
``monopole" could be approximated by the tip of a long thin solenoid.

Extensive numerical studies of the Thomson problem show that the
ground state for a small number of particles, typically $M \leq
150$, consists of twelve positive disclinations (the minimum number 
compatible with Euler's theorem) located at the vertices of an icosahedron
\cite{AWRTSDW:97,EG:97}. Recent results have shown that for 
systems as small as $500$ particles, however, configurations with
additional topological defects \cite{Alar,PGM:99} have lower
energies than icosahedral ones.  

These remarkable results for the Thomson problem raise a number of
important questions, such as the mechanism behind the proliferation
of defects, the nature of these unusual low-energy states, the
universality with respect to the underlying particle potential and the 
generalization to more complex situations.

A formalism suitable to address all these questions has been 
proposed recently\cite{BNT:00}.  Disclinations are considered the
fundamental degrees of freedom, interacting according 
to the energy \cite{BNT:00}
\begin{eqnarray}
\label{inv_Lap}
H=E_0 &+& \frac{Y}{2}\int\int d\sigma(x) d\sigma(y) \nonumber \\
 & & \left[ \bigg ( s(x)-K(x)\bigg )\frac{1}{\Delta^2}\bigg
(s(y)-K(y)\bigg) \right] \, ,
\end{eqnarray}
where the integration is over a fixed surface with area element
$d\sigma(x)$ and metric $g_{ij}$, $K$ is the Gaussian curvature, $Y$ 
is the Young modulus in flat space and $s(x)=\sum_{i=1}^N
\frac{\pi}{3}q_i \delta(x,x_i)$ is the disclination density
$\Big ( \delta(x,x_i)=\delta(x-x_i)/\sqrt{det (g_{ij})} \Big
)$. Defects like dislocations or grain boundaries are built from these
elementary disclinations. $E_0$ is the energy corresponding to a
perfect defect-free crystal with no Gaussian curvature; $E_0$ would     
be the ground state energy for a $2$D Wigner crystal of electrons in
the plane. Eq.(\ref{inv_Lap}) restricted to a 2-sphere gives     
\bea
\label{Def_Int}
H=E_0 &+& \frac{\pi Y}{36} R^2 \sum_{i=1}^{N} \sum_{j=1}^N q_i q_j
\chi(\theta^i,\psi^i;\theta^j,\psi^j) \nonumber \\
&+& N E_{core} \ , 
\eea
where $R$ is the radius of the sphere and $\chi$ is a function
of the geodesic distance $\beta_{ij}$ between defects with polar
coordinates $(\theta^i,\psi^i$ ; $\theta^j,\psi^j)$\cite{BNT:00};
$\chi(\beta)= 1 + \int^{\frac{1-cos\beta}{2}}_0 dz\,\frac{{\rm ln z}}{1-z}$.
Here 5- and  7-fold defects correspond to $q_i=+1$ and $-1$
respectively. In this letter we show that the continuum formalism
embodied in Eq.(\ref{Def_Int}) implies: (a) flat space results for
elastic constants can be bootstrapped into very accurate quantitative 
calculations for generalized Thomson problems, thus providing a 
stringent test of the validity of this approach; (b) new results for
finite length grain boundaries, consisting of dislocations with
variable spacing, in the context of the $2\pi$ disclinations appearing
in tilted liquid crystal phases\cite{DPM:86}; and (c) sufficient
power and generality to determine the ground state for the $8$ minimal
disclinations arising in {\em square} tilings of a sphere. 

The Young modulus appearing in Eq.(\ref{Def_Int}) and the energy $E_0$ may be 
computed in flat space via the Ewald method \cite{BM:77,FHM:79}. The
result for $M$ particles with long range pairwise interactions given
by $e^2/r^{\gamma}$ ($0 < \gamma < 2$) is \cite{BCNT:02}    
\begin{eqnarray}
\label{Young_mod}
Y &=& 4\eta(\gamma)\frac{e^2}{A_C^{1+\gamma/2}} \ , \nonumber \\
\frac{E_0}{Me^2} &=& \theta(\gamma) \left (\frac{4\pi}{A_C}\right )^{\gamma/2}+
\frac{\pi}{A_CR^{\gamma-2}} \rho(\gamma) \ ,
\end{eqnarray}
where $\eta$, $\theta$ and $\rho$ are potential-dependent coefficients
whose numerical values will be reported elsewhere \cite{BCNT:02} and  
$A_C$ is the area per particle. For $M$ particles crystallizing on 
the sphere, $A_C=4\pi R^{2}/M$, and combining Eqs.(\ref{Def_Int}) 
and (\ref{Young_mod}) gives a large $M$ expansion for the ground state
energy,     
\be\label{Energy_ofM}
E_G=\frac{e^2}{2R^{\gamma}}\left[ a_0(\gamma)M^2
-a_{1}(\gamma)M^{1+\frac{\gamma}{2}} + a_2(\gamma)M^{\frac{\gamma}{2}} 
+ \cdots\right] \ ,
\ee
where $a_0(\gamma) = 2^{1 - \gamma}/(2-\gamma)$ and the
subleading coefficients $a_{i}(\gamma)$ depend explicitly
on the potential and on the positions and number 
of disclinations. The first (non-extensive) term is proportional to
$M^2$ and is usually canceled by a uniform background charge for
Wigner crystals of electrons.  The coefficient $a_1$ is
a universal function of the positions of the defects, up to a
potential-dependent constant. Using the results in \cite{BNT:00}, 
theoretical predictions for large $M$ for icosadeltahedral lattices
\cite{Virus:62} of type $(n,0)$ and $(n,n)$ are given in Table
\ref{Tab__Thomson2} for five values of $\gamma$. 

\begin{figure}
\includegraphics[width=3.6in]{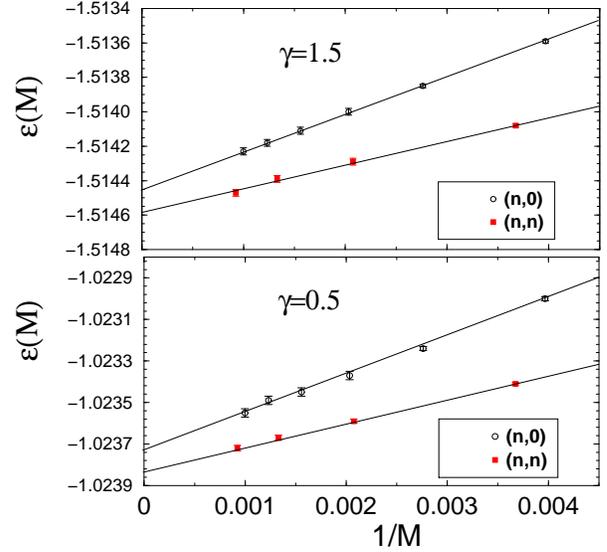}
\caption{Numerical estimate of $\varepsilon (M)$ as a function of
$1/M$ for $(n,0)$ and $(n,n)$ icosadeltahedral lattices with $\gamma 
=(1.5,0.5)$.} 
\label{Both}
\end{figure}

\begin{table}
\centerline{
\begin{tabular}{||c||c|c|c|c||}
\multicolumn{1}{c}{$\gamma$} & \multicolumn{1}{c}{$a_1(\gamma)$} &
\multicolumn{1}{c}{\hspace{.4cm} } &\multicolumn{1}{c}{$(n,n)$} & \multicolumn{1}{c}{$(n,0)$} \\\hline
$1.5$    &  $1.51473$ & &$1.51454(2)$  &  $1.51445(2)$ \\\hline
$1.25$   &  $1.22617$ & &$1.22599(7)$  &  $1.22589(7)$   \\\hline
$1.0$    &  $1.10494$ & &$1.10482(3)$  &  $1.10464(3)$   \\\hline
$0.75$   &  $1.04940$ & &$1.04921(6)$  &  $1.04910(6)$ \\\hline
$0.5$    &  $1.02392$ & &$1.02390(4)$  &  $1.02372(4)$   \\\hline
\end{tabular}}
\vspace{0.1cm}
\caption{Analytical predictions (first column) for a large number of 
particles and values extrapolated from numerical simulations 
(second and third columns) of the coefficient $a_1(\gamma)$, as 
defined in Eq.(\ref{Energy_ofM}) for $(n,n)$ and $(n,0)$
icosadeltahedral lattices. Similar accuracy holds for other values of $\gamma$.} 
\label{Tab__Thomson2}
\end{table}
These predictions may now be compared with direct minimizations of particles
on the sphere in icosadeltahedral configurations by 
fitting the results to Eq.(\ref{Energy_ofM}). 
In Fig.\ref{Both} we plot $\varepsilon (M)$ versus $1/M$ for
$(n,0)$ and $(n,n)$ icosadeltahedral configurations for
$\gamma=1.5$ and $\gamma=0.5$, where
\be
\varepsilon(M)=\frac{[2R^{\gamma}E_G/e^2-a_0(\gamma)M^2]}
{M^{1+\gamma/2}} \ .
\ee
The coefficient $a_1(\gamma)$ is determined by the intercept in the
$M\rightarrow\infty$ limit ($M = 10n^2 + 2$ for $(n,0)$ lattices and
$M=30n^2 +2$ for $(n,n)$ lattices \cite{Virus:62}).    

\begin{figure}
\includegraphics[width=2.9in]{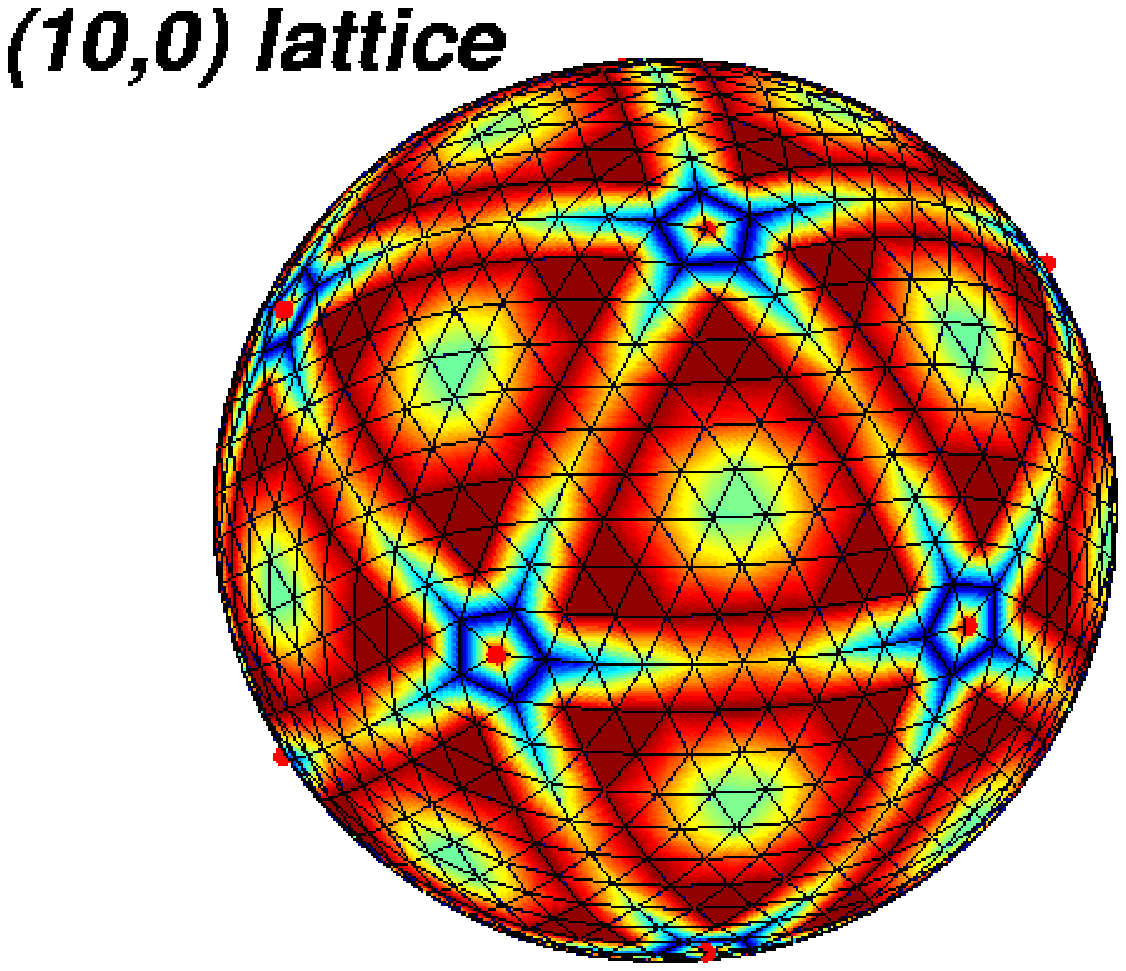}
\includegraphics[width=2.9in]{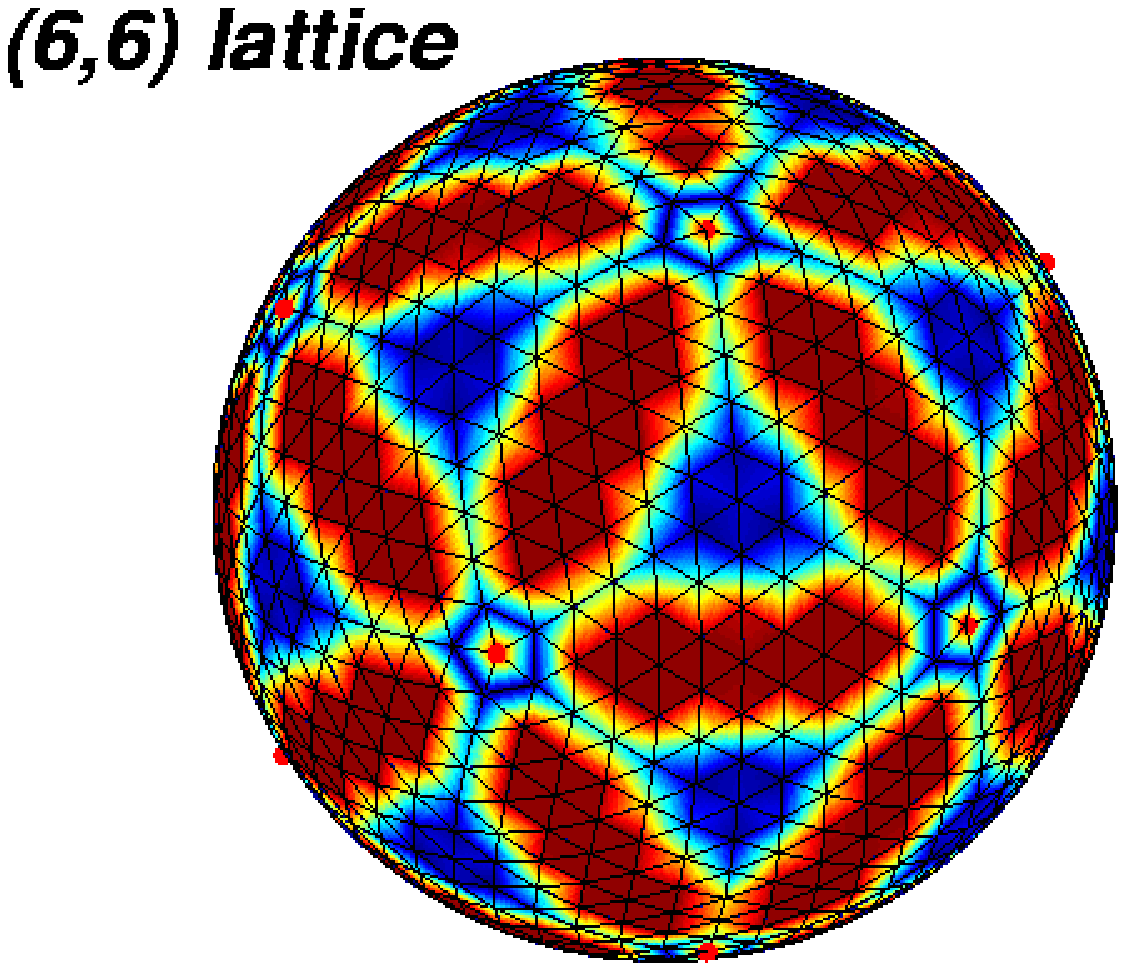}
\caption{Strain energy distribution (red/high, blue/low) for a
(10,0) and a (6,6) configuration.} 
\label{fig__colornn}
\end{figure}
The continuum elastic interaction between disclinations in
Eq.(\ref{Def_Int}) is essential to obtain the correct limiting 
behavior since it reflects contributions both from the energy per
particle in flat space as well as the energies of 12 isolated
disclinations. The defect core energy term in Eq.(\ref{Def_Int})
contributes to the leading correction $a_2(\gamma)$. We find agreement
to four significant figures for $(n,0)$ icosadeltahedral lattices and
to five significant figures for $(n,n)$ lattices. The small residual
difference in energy for $(n,0)$ and $(n,n)$ configurations may be
understood from the differing strain energies shown in
Fig.\ref{fig__colornn}. This small discrepancy may be attributable to
a line tension associated with ridges of minimum or maximum strain
connecting disclinations \cite{BCNT:02}.  

\begin{figure}
\includegraphics[width=2.1in]{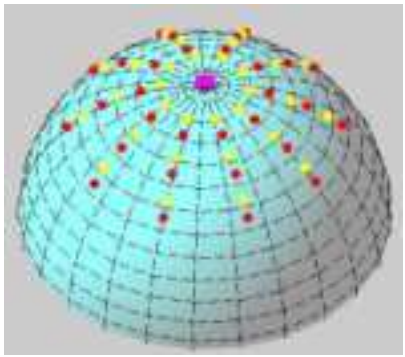}
\caption{A 10-arm grain boundary array emerging from a $+6$ defect
(purple square) at the north pole. Charge $+1$ disclinations are
red and charge $-1$ disclinations are yellow.} 
\label{capstar}
\end{figure}

We now turn to the study of grain boundaries on a sphere using the
model described by Eq.(\ref{Def_Int}). We illustrate the method 
\cite{BNT:00} by considering just two $2\pi$ defects (appropriate to 
crystals of tilted molecules) with suitable boundary conditions 
\cite{DPM:86}. To approximate the $2\pi$ disclination of tilted
molecules in a hemispherical crystal \cite{DPM:86} we replace the
icosahedral configuration of twelve disclinations with two clusters of
six $2\pi/6$ defects at the north and south poles. For simplicity, we
use isotropic elastic theory and neglect nonuniversal details near the
core of the $+2\pi$ disclination. Upon adding just one dislocation
of Burgers vector $b$ (i.e. a $+/- \, 2\pi/6$ disclination pair separated
by distance $b$), the minimum energy in Eq.(\ref{Def_Int}) is 
achieved by a polar angle $\theta_0(b)$ with $\vec{b}$ perpendicular
to the geodesic joining the north and south poles.   
For small numbers of dislocations, the minimum energy  
configuration consists of two polar rings of dislocations located at 
angles $\theta_0(b)$ and $\pi - \theta_0(b)$ relative to the north
pole. The dislocations eventually organize into grain boundaries
centered on $\theta_0(b)$, as shown in Fig.\ref{capstar}. Remarkably,
no other minima were found. 

Because the global minimization just described becomes computationally
demanding for more than thirty defects, further minimizations focused
on a reduced parameter space specified by the orientations and
distances of the grain boundaries from the two $+6$ defects.  
Following \cite{BNT:00}, the system {\em dynamically} chooses the
average lattice spacing $a \equiv b$ that best accommodates the array
structure by extremizing the energy of Eq.(\ref{Def_Int}).

For reasonable parameter values in our continuum description, both two
antipodal $+6$  disclinations and the icosadeltahedral configurations
are indeed unstable to the formation of grain boundaries for
sufficiently large sphere radius\cite{Footnote,Alar,PGM:99,BNT:00}.    
In a flat monolayer, the spacing between dislocations in $m$-grain 
boundary arms radiating from a disclination of charge $s$ is given by
\cite{HiLo:68} 
\be\label{space_disl}
l=\frac{b}{2 \sin \left ( \frac{s}{2m} \right )}  \ ,
\ee
where $b$ is the Burgers vector charge. The disclination charge is 
$s=\frac{2 \pi}{6} p$ ($p=1$ or $6$, corresponding to the Thomson 
problem or tilted molecules, respectively). To generalize this result
for symmetrical grain boundaries on a sphere, consider the Burgers
circuit formed by an isosceles spherical triangle with apex angle
$\phi_{gb}=s/m$ at a disclination at the north pole and centered on
one of the $m$-grain boundary arms \cite{CN:93}. If the altitude of
this triangle is $h$, the net Burgers vector $B$ defined by this
circuit is given by the geodesic distance spanning the base of this
triangle. A straightforward exercise in spherical trigonometry leads
to 
\be\label{Sph_Total}
\cos \left ( \frac{B}{R} \right ) = \frac{ \cot^2(h/R) \cos^2(\phi_{gb}/2)+\cos(\phi_{gb})}
{1+\cot^2(h/R)\cos^2(\phi_{gb}/2)} \ .
\ee
Writing $B \approx b\int^h_0 dh'/l(h')$, where $l(h)$ is a (variable)
dislocation spacing \cite{BNT:00}, we can invert this formula and thus
generalize Eq.(\ref{space_disl}) to the sphere.

\begin{figure}[ctb]
\includegraphics[width=2.1in]{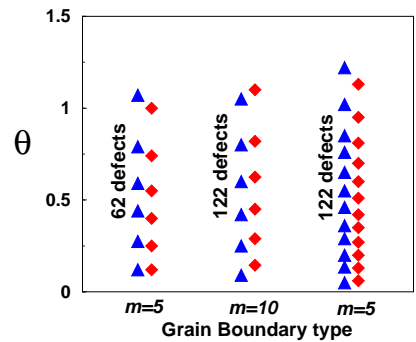}
\caption{Defect positions obtained from minimization (up triangle)  
and from Eq.(\ref{Sph_Total}) (diamonds).} 
\label{fig__distance_var_spac}
\end{figure}

Results comparing our minimization with Eq.(\ref{Sph_Total}) are shown
in Fig.\ref{fig__distance_var_spac}. Both approaches predict the same
number of dislocations within a grain and dislocation spacings which
{\em increase} with $\theta$ \cite{BNT:00}. The small discrepancies in 
the positions of the dislocations are presumably due to interactions
between dislocations in different arms. 

The physics associated with Eq.(\ref{Def_Int}) is remarkably general.  
We sketch here how the results above may be adapted to a sphere 
tiled with a {\em square} lattice.  A planar square lattice 
is described by three elastic constants (as opposed to the two
Lam\'{e} coefficients in the triangular case), leading to an energy 

\be\label{LL_SQ}
H=\frac{\lambda_{\alpha \beta,\mu \nu}}{2}\int d^2{\bf x} \, u_{\alpha \beta}
u_{\mu \nu} \ ,
\ee
where the independent elastic constants are
$\lambda_{11,11},\lambda_{11,22}$ and $\lambda_{12,12}$.  
The same derivation as that leading to Eq.(\ref{inv_Lap}) now leads to
\be\label{sq_en}
H=\frac{1}{2}\int d\sigma(x) d\sigma(y) 
\Big (K(x)-s(x) \Big ) G(x,y)
\Big (K(y)-s(y) \Big ) \ ,
\ee
where $G(x,y)=(\frac{1}{Y}\Delta^2+2\epsilon
\nabla^2_1\nabla^2_2)^{-1}$, with $\nabla_{\rm i}$ the gradient 
in direction ${\rm i}=1$ or $2$, as defined by the local square
lattice. The fundamental defects are now $\pm \frac{\pi}{2}$
disclinations. The elastic constants are    
\begin{eqnarray}\label{el_cons}
Y &=& \frac{\lambda^2_{11,11}-\lambda^2_{11,22}}{\lambda_{11,11}} \quad , \nonumber \\
\epsilon &=& -\frac{\lambda_{11,22}+\lambda_{11,11}}{\lambda^2_{11,11}-\lambda^2_{11,22}}
+\frac{1}{2 \lambda_{12,12}} \ .
\end{eqnarray}
The interaction energy for disclinations in a square lattice becomes   
equivalent to Eq.(\ref{inv_Lap}) only in the limit $\epsilon=0$. 
Although details such as the critical value of $R/a$ for 
the onset of grain boundaries will differ, we expect that the physics
remains essentially the same. For small numbers of
particles the ground state will consist of 8 $q=+1$ disclinations. In 
the isotropic case ($\epsilon=0$) the ground state is a distorted
cube, with one face twisted by $45^{\circ}$\cite{BCNT:02}, similar to
tetratic liquid crystal ground states on the sphere \cite{LubPro:92}. 

We expect similar results for geometries other than the sphere. 
For isotropic crystals on a torus with the right aspect ratio, for
example, one might expect 12 5-fold disclinations on the outer wall
(where the Gaussian curvature is positive) and twelve compensating
7-fold disclinations on the inner wall (where the Gaussian curvature
is negative) to play the role of the icosahedral configurations on the
sphere. As more particles are placed on the torus, we expect grain
boundaries to emerge from these disclinations. 

There are some important issues left for a future publication
\cite{BCNT:02}, such as a more detailed derivation of the asymptotic
expansion Eq.(\ref{Energy_ofM}) and a reliable determination of the 
optimal number of arms within a grain boundary. We hope the calculations
presented here will convince the reader of the usefulness of our
approach.  

It is a pleasure to thank Alar Toomre for sharing with us his
numerical work on the original Thomson problem. The work of M.B. and
A.C. has been supported by the U.S. Department of Energy (DOE) under
Contract No. DE FG02 85ER40237. The research of D.R.N was supported by
the National Science Foundation through Grant No. DMR97-14725 and
through the Harvard Material Research Science and Engineering
Laboratory via Grant No. DMR98-09363. The work of A.T. has been
supported by the Materials Computation Center and grants NSF-DMR
99-76550 and NSF-DMR 00-72783.


\vspace{-0.1cm}

\begin{thebibliography}{10}

\bibitem{Thom:04}
J.~J. Thomson, Philos.\ Mag. {\bf 7}, 237 (1904).

\bibitem{Sku:97}
E.B. Saff and A.B.J. Kuijlaars, Math. Intelligencer {\bf 19}, 5 (1997).

\bibitem{Lei:93}
P.~Leiderer,
Z. Phys. {\bf B98}, 303 (1993).


\bibitem{Dav:97}
E.J. Davis, Aerosol Sci. Techn. {\bf 26}, 212 (1997).

\bibitem{Virus:62}
D.L.D. Caspar and A.~Klug,
Cold Spring Harbor Symp. Quant. Biol. {\bf 27}, 1 (1962).

\bibitem{Virus:93}
C.J. Marzec and L.A. Day,
Biophys. Jour. {\bf 65}, 2559 (1993).

\bibitem{Nik:01}
M.~Nikolaides,
Thesis, Physics Department, TU Munich, 2001.

\bibitem{DHNMBW:02}
A.D. Dinsmore, M.F. Hsu, M. Nikolaides, M. Marquez, A.R. Bausch and
D.A. Weitz, {\em Colloidosomes: Selectively-Permeable Capsules
Composed of Colloidal Particles} (to be published). 

\bibitem{Kr:85}
H.W.~Kroto et~al, Nature {\bf 318}, 162 (1985).

\bibitem{Rings:2001}
J.O.M. Sano, A. Kamino, and S. Shinkai,
Science {\bf 293}, 1299 (2001).

\bibitem{DoMo:97}
M.J.W. Dodgson and M.A. Moore,
Phys. Rev. B {\bf 55}, 3816 (1997)
(arXiv:cond-mat/9512123). 

\bibitem{AWRTSDW:97}
E.L. Altschuler, T.J. Williams, E.R. Ratner, R.~Tipton, R.~Stong, F. Dowla and
F.~Wooten, Phys. Rev. Lett. {\bf 78}, 2681 (1997).

\bibitem{EG:97}
T.~Erber and G.M. Hockney,
Adv. Chem. Phys. {\bf 98}, 495 (1997).

\bibitem{Alar}
A.~Toomre,
private communication.

\bibitem{PGM:99}
A.~Perez-Garrido and M.~A. Moore,
Phys. Rev. B {\bf 60}, 15628 (1999) 
(arXiv:cond-mat/9905217) and references therein.

\bibitem{BNT:00}
M.~Bowick, D.~R. Nelson, and A.~Travesset,
Phys. Rev. B {\bf 62}, 8738 (2000)
(arXiv:cond-mat/9911379).


\bibitem{DPM:86}
R.~Pindak, S.B. Dierker, and R.B. Meyer,
Phys. Rev. Lett. {\bf 56}, 1819 (1986); by introducing a 
pressure difference across the free standing liquid crystal film, one
could study the $2\pi$ disclinations and grain boundaries of this
reference in a hemispherical environment (R. Pindak and C.C. Huang,
private communication). 

\bibitem{BM:77}
L.~Bonsall and A.A. Maradudin,
Phys. Rev. B {\bf 15}, 1959 (1977).

\bibitem{FHM:79}
D.~Fisher, B.~Halperin, and R.~Morf,
Phys. Rev. B {\bf 20}, 4692 (1979).

\bibitem{BCNT:02}
M.~Bowick, A.~Cacciuto, D.R. Nelson, and A.~Travesset,
in preparation.

\bibitem{HiLo:68}
J.~P. Hirth and J.~Lothe,
{\em Theory of Dislocations} (Wiley, New York, 1982).

\bibitem{CN:93}
C.~Carraro and D.~R. Nelson,
Phys. Rev. E {\bf 48}, 3082 (1993)
(arXiv:cond-mat/9307008).

\bibitem{Footnote}
For the purposes of determining grain boundary stability it is useful
to write the core energy term $E_{core}$ in Eq.(\ref{Def_Int}) as $N
E_{core} = 12E_5 + \frac{N-12}{2}E_d$ and parametrize the physics in
terms of a {\em dislocation} core energy $E_d$ of order $0.1
Ya^2$. For antipodal $+6$ defects, $NE_{core} = 2E_{+6} +
\frac{N-2}{2}E_d$.  

\bibitem{LubPro:92}
T.C. Lubensky and J. Prost, 
J. Phys. II {\bf 2}, 371 (1992).

\end{thebibliography}
\end{document}